# Mössbauer study of the '11' iron-based superconductors parent compound $Fe_{1+x}Te$

A. Błachowski[1], K. Ruebenbauer[1*], P. Zajdel[2], E. E. Rodriguez[3], and M. A. Green[3,4]

[1]Mössbauer Spectroscopy Laboratory, Pedagogical University
PL-30-084 Kraków, ul. Podchorążych 2, Poland

[2]Division of Physics of Crystals, Institute of Physics, Silesian University
PL-40-007 Katowice, ul. Uniwersytecka 4, Poland

[3]NIST Center for Neutron Research, NIST
100 Bureau Dr., Gaithersburg, MD 20878, U.S.A.

[4]Department of Materials Science and Engineering, University of Maryland
College Park, MD 20742, U.S.A.

[*]Corresponding author: sfrueben@cyf-kr.edu.pl

**PACS:** 74.70.Xa, 75.30.Fv, 76.80.+y

**Keywords:** iron-based superconductors, spin density waves, Mössbauer spectroscopy

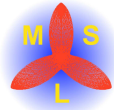

## Abstract

$^{57}Fe$ Mössbauer spectroscopy was applied to investigate the superconductor parent compound $Fe_{1+x}Te$ for x=0.06, 0.10, 0.14, 0.18 within the temperature range 4.2 K – 300 K. A spin density wave (SDW) within the iron atoms occupying regular tetrahedral sites was observed with the square root of the mean square amplitude at 4.2 K varying between 9.7 T and 15.7 T with increasing x. Three additional magnetic spectral components appeared due to the interstitial iron distributed over available sites between the Fe-Te layers. The excess iron showed hyperfine fields at approximately 16 T, 21 T and 49 T for three respective components at 4.2 K. The component with a large field of 49 T indicated the presence of isolated iron atoms with large localized magnetic moment in interstitial positions. Magnetic ordering of the interstitial iron disappeared in accordance with the fallout of the SDW with the increasing temperature.

## 1. Introduction

The compound $Fe_{1+x}Te$ is a parent compound of the '11' iron-based superconductors and can be prepared by conventional solid-state techniques with excess iron x only in the range x=0.04 – 0.18 [1, 2]. Superconductivity at a critical temperature of 15 K under ambient pressure is induced by partial substitution of tellurium by selenium [3, 4] or sulfur [5, 6]. $Fe_{1+x}Te$ crystallizes in the tetragonal structure in the magnetically disordered state and becomes monoclinic upon magnetic ordering for x < 0.12 or orthorhombic for higher amount of interstitial iron with the mixed phase region extending between x=0.11 - 0.12 [7, 2, 8, 9]. In contrast orthorhombic FeSe is a superconductor under ambient pressure below 8.5 K without any magnetic moment on the iron site [10, 11]. For $Fe_{1+x}Te$ the magnetic ordering is complex as a result of competing interactions. Iron on the regular tetrahedral sites develops bicollinear antiferromagnetic structure for the almost stoichiometric compound, transforming into an incommensurate spin density wave (SDW) upon increasing x [7, 2, 9]. The transverse SDW propagates along the a-axis with the moment pointing along the b-direction [2, 12]. Above x ≥ 0.12 some magnetic moment along the c-axis appears and the structure is interpreted as developing elliptical helicity in the b-c plane tending towards circular helicity for the highest iron content [2, 9]. The excess iron is located between regular Fe-Te layers occupying partly available interstitials [13]. It is interesting to look upon magnetic behavior of the excess iron. Density functional calculations suggest the presence of a significant magnetic moment on the interstitial iron atoms [14].

The compound $Fe_{1+x}Te$ has been previously investigated by means of the $^{57}Fe$ Mössbauer spectroscopy [15-17] and very complex magnetic spectra were found. The aim of this study is to understand how the amount of the interstitial iron affects the magnetic properties of the $Fe_{1+x}Te$ through temperature dependent $^{57}Fe$ Mössbauer spectroscopy.

## 2. Experimental

Samples of $Fe_{1+x}Te$ with x=0.06(1), 0.10(1), 0.14(1) were prepared as single crystals by the Bridgman method, while the sample with x=0.18(1) was prepared in powder form. The chemical compositions were determined from a combination of neutron diffraction data and X-ray fluorescence. Further details of the sample preparation and determination of the compositions have been previously described in Refs [2, 13]. The Mössbauer spectra were collected on the powdered samples mixed with the $B_4C$ carrier and absorbers were made of about 25 mg/cm$^2$ of the investigated material. Low velocity spectra for x=0.10, 0.18 were collected for samples having 30 mg/cm$^2$. A commercial $^{57}Co(Rh)$ source kept at room temperature was used and all shifts are reported versus room temperature α-Fe. The MsAa-3 spectrometer was used with the Kr-filled proportional counter to collect 14.41-keV Mössbauer spectra of $^{57}Fe$. The velocity scale was calibrated by the laser equipped Michelson-Morley interferometer. The sample temperature was maintained with the help of the Janis Research Co. Inc. cryostat SVT-400M. Spectra were processed by the *GmfpHARM* application belonging to the *MOSGRAF-2009* suite [18, 19].

## 3. Results and discussion

The Mössbauer results for four samples are shown in Figures 1-4. The first column of each figure contains $^{57}Fe$ Mössbauer spectra versus temperature. Spectra are labeled on the left side by parameters of the SDW component including contribution of the SDW to the cross-section in percent, root mean square (RMS) amplitude of the SDW $\sqrt{\langle B^2 \rangle}$ in Tesla [18], total shift $S$ versus α-Fe, and effective quadrupole splitting Δ. Spectra are labeled on the right side by contributions of the non-magnetic component (NM) and excess iron components. For the excess iron components their respective magnetic hyperfine fields are shown. Additionally, Figure 4 shows contributions and hyperfine fields due to unreacted iron (~33 T) and resulting oxide (~50 T). The oxide component is masked below 60 K by the component due to the excess iron with the highest hyperfine field and very broad lines. The top row of each figure shows room temperature (RT) and 79 K spectra measured on the low velocity scale. The second column shows distributions of the magnetic hyperfine fields generated by the SDW with the mean field of the distribution <B>. The third column shows the shape of the SDW with the maximum amplitude $B_{max}$ and amplitudes of the first two dominant harmonics $h_n$ [18]. Errors of the quantities displayed in Figures 1-4 are of the order of unity for the last digit shown. Spectra were processed within the transmission integral approximation with several components describing various iron states.

Parameters of the $Fe_{1+x}Te$ room temperature spectra are given in Table I. Iron in the Fe-Te layers shows up as doublet having shift of about 0.5 mm/s and splitting of 0.3 mm/s. Excess iron for samples with x=0.06, 0.10 shows up as a singlet and doublet with large splitting. The singlet splits into a doublet for x=0.14. For x=0.18 one cannot distinguish between various states of the excess iron, and it acquires parameters similar to those of the iron in the Fe-Te layers. For sample with x=0.18 some contribution due to unreacted α-Fe and resulting magnetically ordered at room temperature $Fe^{3+}$ (high spin) oxide is observed.

**Table I**

Parameters of the $Fe_{1+x}Te$ room temperature spectra. Symbols have the following meaning: *A* - relative contribution to the cross-section, *S* – spectral shift, Δ – quadrupole splitting, Γ – absorber line-width. The absorber width is common for all components except for x=0.14, where the larger value shows up for the third component. Errors are of the order of unity for the last digit shown.

| $Fe_{1+x}Te$ x | $A_1$ (%) | $S_1$ (mm/s) | $\Delta_1$ (mm/s) | $A_2$ (%) | $S_2$ (mm/s) | $\Delta_2$ (mm/s) | $A_3$ (%) | $S_3$ (mm/s) | $\Delta_3$ (mm/s) | Γ (mm/s) |
|---|---|---|---|---|---|---|---|---|---|---|
| 0.06 | 84 | 0.495 | 0.315 | 12 | 0.28 | 0 | 4 | 0.22 | 0.84 | 0.21 |
| 0.10 | 87 | 0.462 | 0.310 | 6 | 0.29 | 0 | 7 | 0.36 | 1.03 | 0.27 |
| 0.14 | 86 | 0.481 | 0.330 | 6 | 0.23 | 0.23 | 9 | 0.34 | 0.96 | 0.29/0.53 |
| 0.18 | 77 | 0.492 | 0.295 | 23 | 0.47 | 0.57 | | | | 0.25 |

Low temperature spectra exhibit complex magnetic structure. The main component is due to iron on the regular tetrahedral sites and it could be described by an incommensurate SDW containing several subsequent odd harmonics varying in number from 3 to 8. A quadrupole splitting of the first excited $^{57}Fe$ state has been accounted for in the first order approximation. One has additional three magnetically split components due to the excess iron. They are described by separate magnetic hyperfine fields and small electric quadrupole interaction accounted for in the first order approximation. Close to the magnetic transition one has to include extra non-magnetic NM component described by the quadrupole split doublet having 0.6 mm/s shift, 0.4 mm/s splitting and 0.2 mm/s line-width.

The SDW has almost rectangular shape close to saturation and evolves with increasing temperature in such a manner that significant parts of the sample have a small hyperfine field. The shape of SDW becomes more irregular with increasing concentration of the excess iron which also decreases the magnetic ordering temperature. On the other hand, RMS amplitude of the SDW at 4.2 K is almost constant until x=0.10 and increases for higher excess iron concentrations. SDW amplitude at low temperature is particularly enhanced close to the highest possible concentration of the excess iron. The total spectral shift amounts to 0.58 mm/s for samples with x=0.06, 0.10 and 0.63 mm/s for x=0.14, 0.18 at 4.2 K. Hence, the electron density on the regular iron site is lower for the orthorhombic structure in comparison with the monoclinic structure by 0.17 el./a.u.$^3$ [20]. The effective quadrupole splitting at 4.2 K is similar and positive ($\Delta$=+0.12 mm/s) for the lowest values of x=0.06, 0.10 (monoclinic), but becomes larger (+0.15 mm/s) for x=0.14 (orthorhombic) and finally it changes the sign and value (-0.04 mm/s) for x=0.18. The last change is likely to be due to the development of the more complex SDW than planar in accordance with the neutron scattering data [2]. Hence, the complication of SDW appears much above the structural change and it is likely to be due to the increasing excess iron content. The quadrupole splitting evolves in the SDW order region for sample with x=0.14 from about +0.03 mm/s at 50 K till +0.15 mm/s at 4.2 K, while for other compositions variation is much smaller across the temperature region of the SDW order. Such behavior is consistent with the phase diagram of Ref. [9] indicating that for x=0.14 the SDW magnetic order occurs at temperature higher than the structural transition.

Additional spectral components exhibit higher magnetic hyperfine fields than those due to the SDW. The magnetic order of these components disappears altogether with the SDW order upon increasing temperature. Hence, these spectral components must originate in the same phase as the phase bearing SDW, and therefore they are due to the excess (interstitial) iron in $Fe_{1+x}Te$.

The excess iron with the highest hyperfine field ranging 48-50 T at 4.2 K is quite unusual for the system investigated. The isomer shift does not differ significantly from the shift of the SDW component staying at *S*=0.5 mm/s and the quadrupole interaction is almost absent. On the other hand, such large field is a strong indication of the high and localized magnetic moment [14]. This spectral component exhibits very large line-widths in comparison with the remaining components even at the lowest temperatures. Line-widths evolve from about 1 mm/s at 4.2 K till about 3 mm/s close to the magnetic transition, while the average field changes little within this temperature range. Therefore it is likely that the localized moment is stabilized by the SDW and one has many local magnetic states separated by small energies. The line is broad close to saturation due to the summation of the local and SDW fields, the latter varying from one atom to another. However, significant broadening with the increasing temperature is probably due to the increasing thermal occupation of the higher energy local magnetic states having sufficiently long lifetimes to contribute to the hyperfine field distribution. The contribution to the absorption cross-section from this component is uncertain for the x=0.18 sample due to the presence of the $Fe^{3+}$ (high spin) magnetically ordered phase resulting from the oxidation of the unreacted iron. However, these two spectral components differ by the line-widths, i.e. the parasitic oxide phase exhibits narrow lines.

The component with an intermediate field ranging 21-24 T at 4.2 K has narrower lines of 0.3 mm/s. The isomer shift of this component at 4.2 K increases from 0.36 mm/s for x=0.06 to about 1 mm/s for x=0.10, i.e. within the monoclinic phase. Afterwards it stays approximately constant at 0.7 mm/s for x=0.14, 0.18 in the orthorhombic phase. A large change of the isomer shift between RT and 4.2 K of about 0.5 mm/s is observed for this component, i.e., much greater than expected due to second order Doppler shift alone. The increase of the shift by 0.5 mm/s on going from RT to the ground state means that the electron density on these iron atoms is significantly reduced with lowered temperature. Hence, some very narrow s-type band must be strongly depopulated by lowering temperature. It is likely that such band is generated by a relatively high concentration of defects, i.e. by the excess iron [21]. The absolute value of the quadrupole splitting is much larger in the monoclinic phase (about 0.6 mm/s) than in the orthorhombic phase (0.2 mm/s).

The lowest field component within the range 15-16 T is best resolved for the x=0.06 sample since the increasing SDW amplitude with the increasing iron concentration makes it more difficult to detect. It is practically impossible to separate this component for the x=0.18 sample. The contribution of this component to the spectral shape is difficult to determine except for x=0.06 due to the overlap with the major SDW component. It is likely that for x=0.10, 0.14 one overestimates contribution of this component because part of the SDW is accounted as the lowest field component due to the overlap. On contrary, for x=0.18 the lowest hyperfine field component is practically accounted for in the SDW. Lines are generally narrow ($\Gamma$=0.5 mm/s) and the isomer shift is very close to the shift of the SDW ($S$=0.6 mm/s). The effective quadrupole splitting is essentially absent for this component.

### Table II

Essential parameters describing evolution of the RMS amplitude of the SDW $\sqrt{\langle B^2 \rangle}$ versus temperature. The symbol $T_c$ is the temperature at which coherent part appears upon cooling. The symbol $T_0$ stands for the temperature at which incoherent part appears upon heating, $B_0$ stands for saturation field, $B_F$ denotes field at bifurcation into coherent and incoherent parts, $\alpha_0$ is a critical exponent below transition, $\gamma$ stands for the parameter describing evolution of the exponent upon cooling to the ground state, and $\beta$ denotes exponent describing evolution of the incoherent part. For details see Ref. [18].

| $Fe_{1+x}Te$ | x = 0.06 | x = 0.10 | x = 0.14 | x = 0.18 |
|---|---|---|---|---|
| $T_c$ (K) | 73(2) | 70(2) | 53.3(4) | 66.0(3) |
| $T_0$ (K) | 69(2) | 67(2) | 51.9(4) | 65.3(3) |
| $B_0$ (T) | 10.10(4) | 9.86(9) | 11.72(4) | 15.74(5) |
| $B_F$ (T) | 7.0(4) | 7.0(7) | 8.3(1) | 9.9(4) |
| $\alpha_0$ | 0.12(1) | 0.10(1) | 0.10(1) | 0.10(1) |
| $\gamma$ | 0.8(2) | 0.6(4) | 0.2(1) | 1.0(1) |
| $\beta$ | 2.2(1) | 2.7(3) | 3.6(1) | 9.4(4) |

The evolution of the RMS amplitude of the SDW $\sqrt{\langle B^2 \rangle}$ versus temperature is shown in Figure 5. Experimental data were fitted within the model described in Ref. [18] and the results are summarized in Table II. The magnetic transition temperature $T_c$ decreases with addition of the excess iron till about x=0.14 and partly recovers for the sample saturated with the excess iron. Hence, the magnetic order is governed by development of the SDW being increasingly perturbed by addition of the interstitial iron. It is likely that randomly distributed localized magnetic moments of the interstitial iron act as the scattering centers for SDW leading to the phase incoherence on relatively short distances. Change from the planar SDW to the more complex helical form [2] at highest concentrations of the interstitial iron seems to enhance exchange forces leading to the partial recovery of the transition temperature. The saturation field $B_0$ remains fairly constant in the monoclinic phase (x=0.06, 0.10) and increases within the orthorhombic phase (x=0.14). Further strong increase is observed with the development of the helicity of SDW (x=0.18). The magnetic hyperfine fields due to SDW behave similarly versus temperature as in other parent compounds of the iron-based superconductors, e.g. in $AFe_2As_2$ (A=Ca, Ba, Eu) [18]

## 4. Conclusions

Despite the existence of a single crystallographic site for the excess iron one observes at least three different kinds of these atoms. Such a situation could occur due to the partial filling of the available interstitial sites by iron and a possibility for some ordering of the iron atoms on these sites. The site with the highest magnetic hyperfine field is likely to contain almost isolated ions, i.e., surrounded by the vacancies on the interstitial sites. Any kind of the order on the interstitial sites is of the short range type as it is invisible by the diffraction methods.

The magnetism of the excess iron and SDW are coupled mutually. The excess iron sees some contribution to the hyperfine field due to SDW, while the SDW shape irregularity is due to the randomly distributed interstitial iron. Both kinds of magnetism disappear at the same transition temperature. Interstitial iron leads to the similar irregularity of the SDW shape as irregularity due to the substitutional dopants in the '122' pnictides [22]. The situation here is completely different from the situation with the 4f magnetic ions located between Fe-As layers. In the case of $EuFe_2As_2$ the 4f and 3d magnetism seem to be almost independent and one can observe coexistence of the 4f magnetism and superconductivity within the same electronic system [23]. On the other hand, such coexistence was not found up to date for 3d magnetism strongly supporting hypothesis that the Cooper pairs are formed from the $s+id$ states [24].

Strong temperature dependence of isomer shift of the excess iron exhibiting intermediate hyperfine field is an indication that the electron density between Fe-Te layers evolves significantly with temperature enhancing two-dimensional character of the material at low temperature.

Interstitial iron has relatively large localized magnetic moment at least for the site with the highest hyperfine field. These moments are almost randomly distributed over the interstitial sublattice. Therefore they interact strongly with the electrons having ability to form Cooper pairs and prevent appearance of superconductivity. One has to remove this iron to have a chance to get superconducting material [25, 26]. Partial replacement of tellurium by either selenium or sulfur removes interstitial iron and one can get superconductor. Some minor components of the selected alcoholic beverages like weak organic acids remove excess iron as well [27, 28].

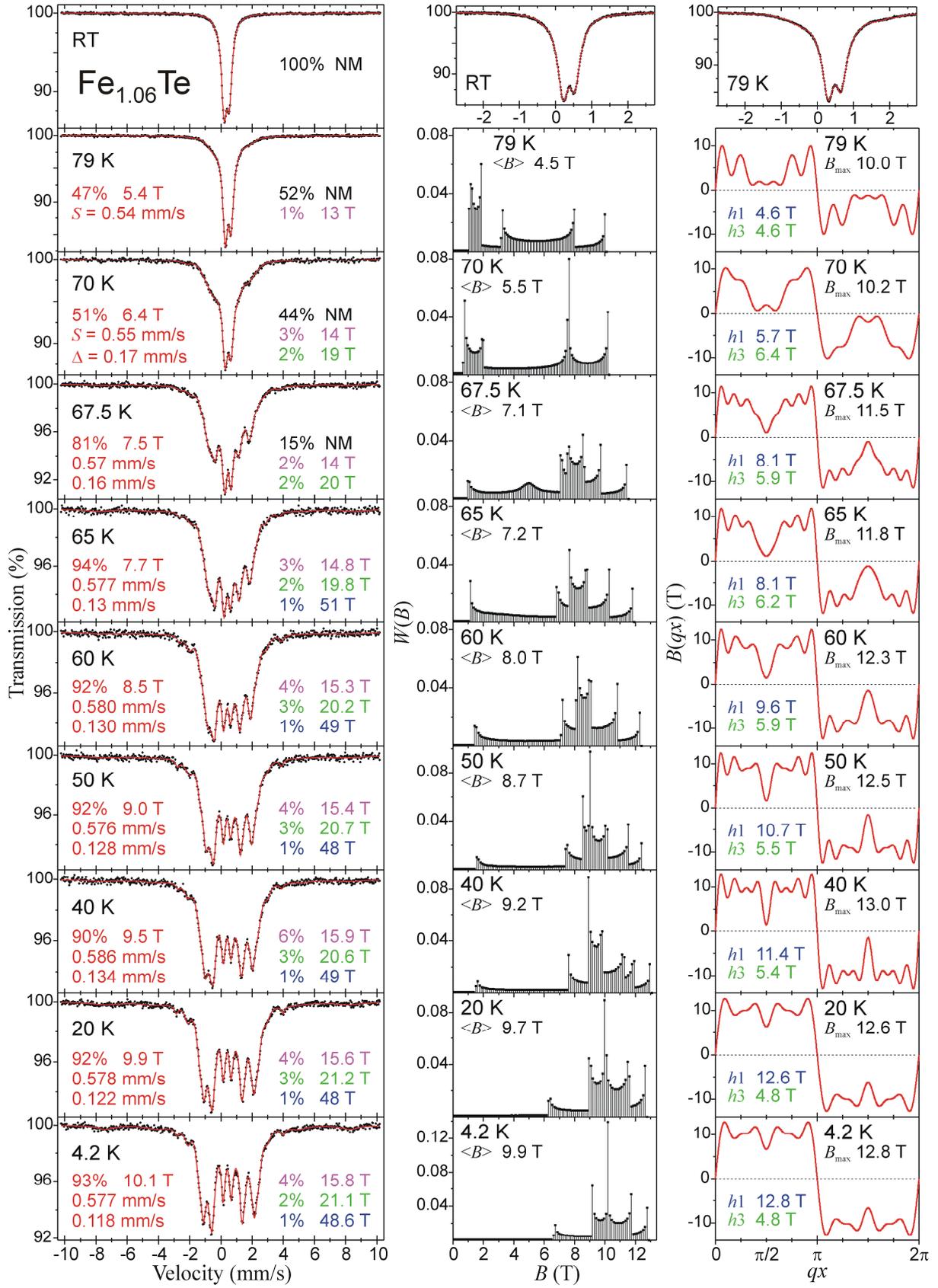

Fig. 1 $^{57}$Fe Mössbauer spectra, hyperfine field distributions of SDW and SDW shape versus temperature for Fe$_{1.06}$Te. See text for details.

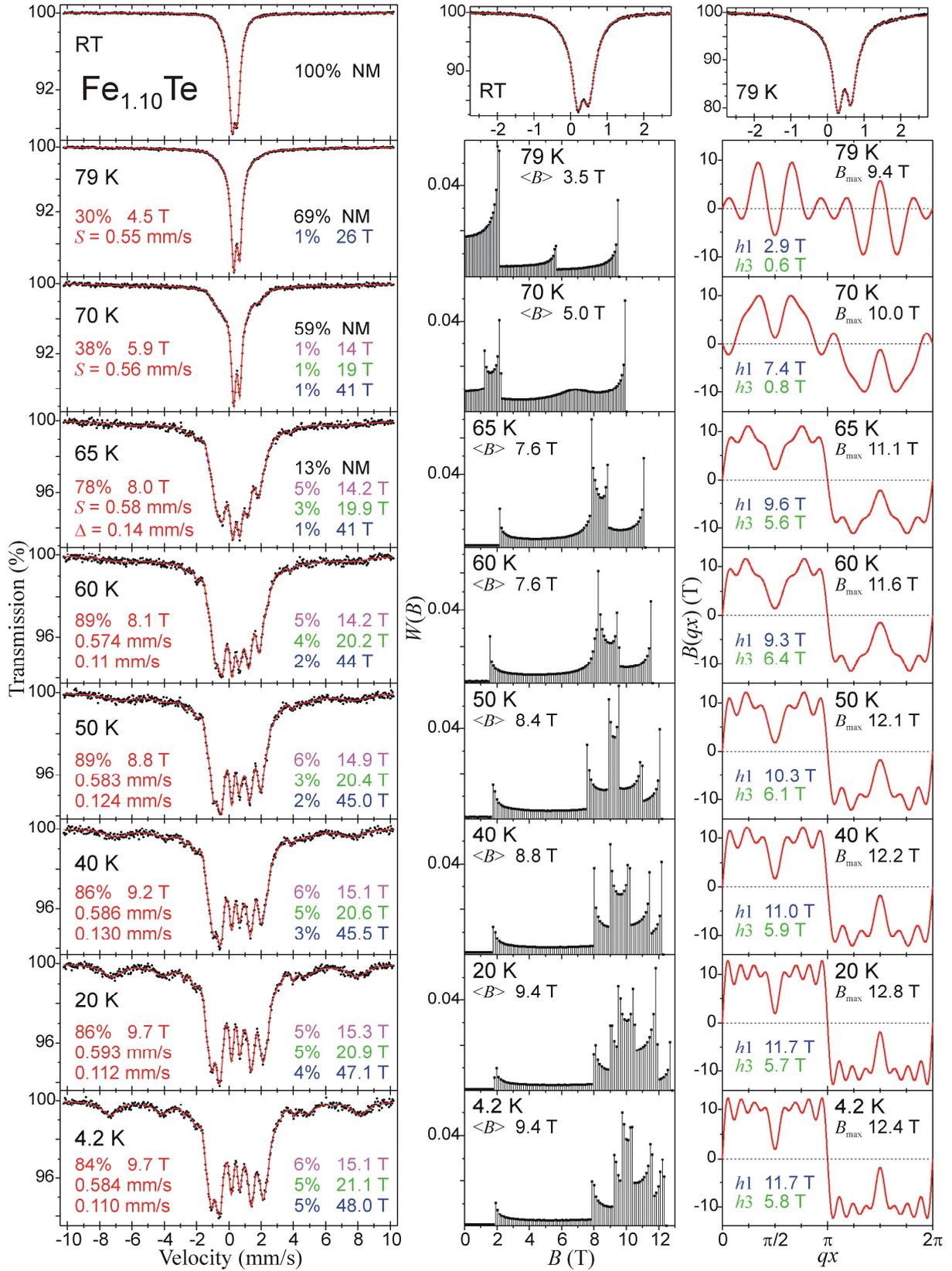

Fig. 2 $^{57}$Fe Mössbauer spectra, hyperfine field distributions of SDW and SDW shape versus temperature for $Fe_{1.10}Te$. See text for details.

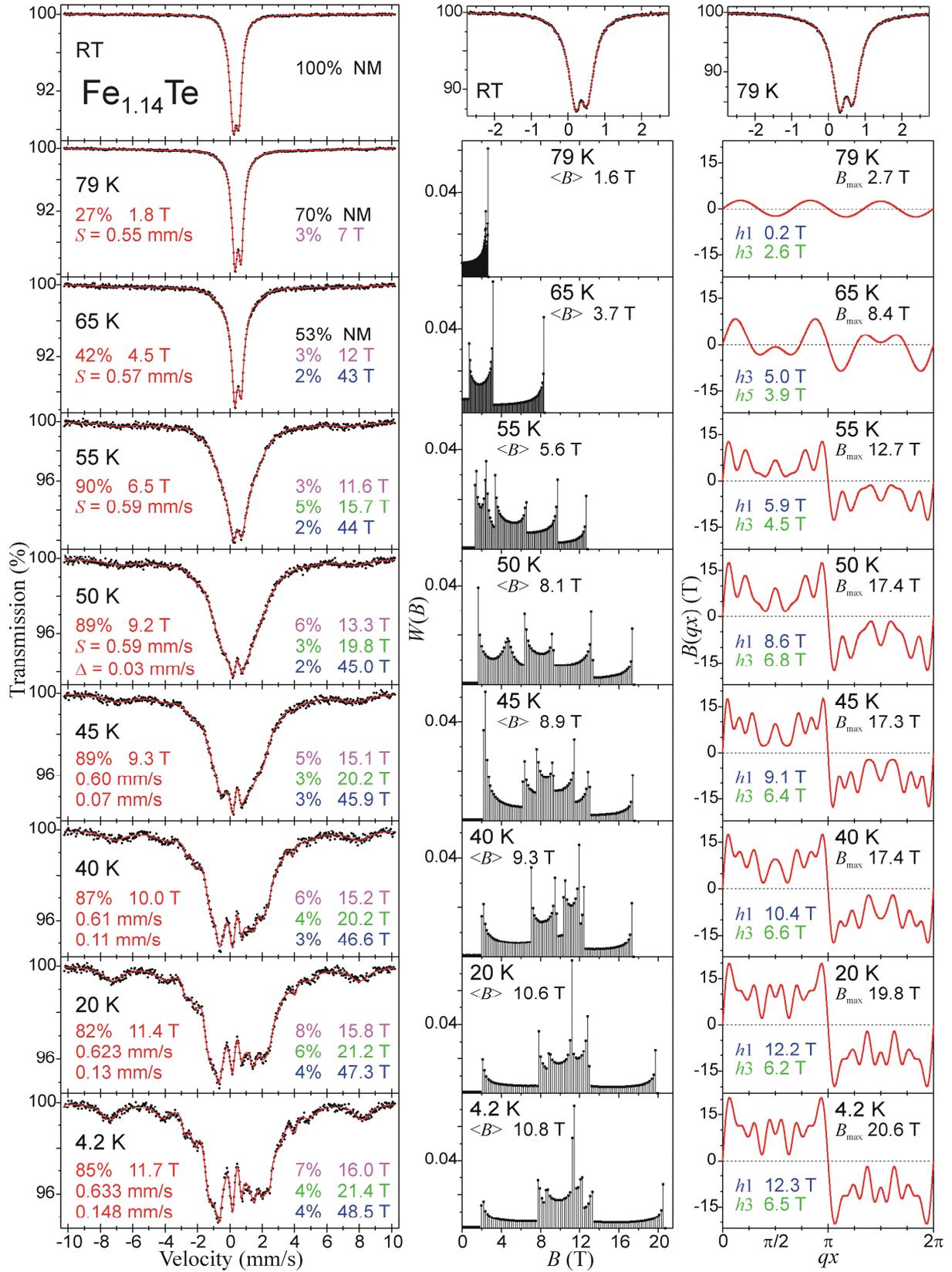

Fig. 3 $^{57}$Fe Mössbauer spectra, hyperfine field distributions of SDW and SDW shape versus temperature for Fe$_{1.14}$Te. See text for details.

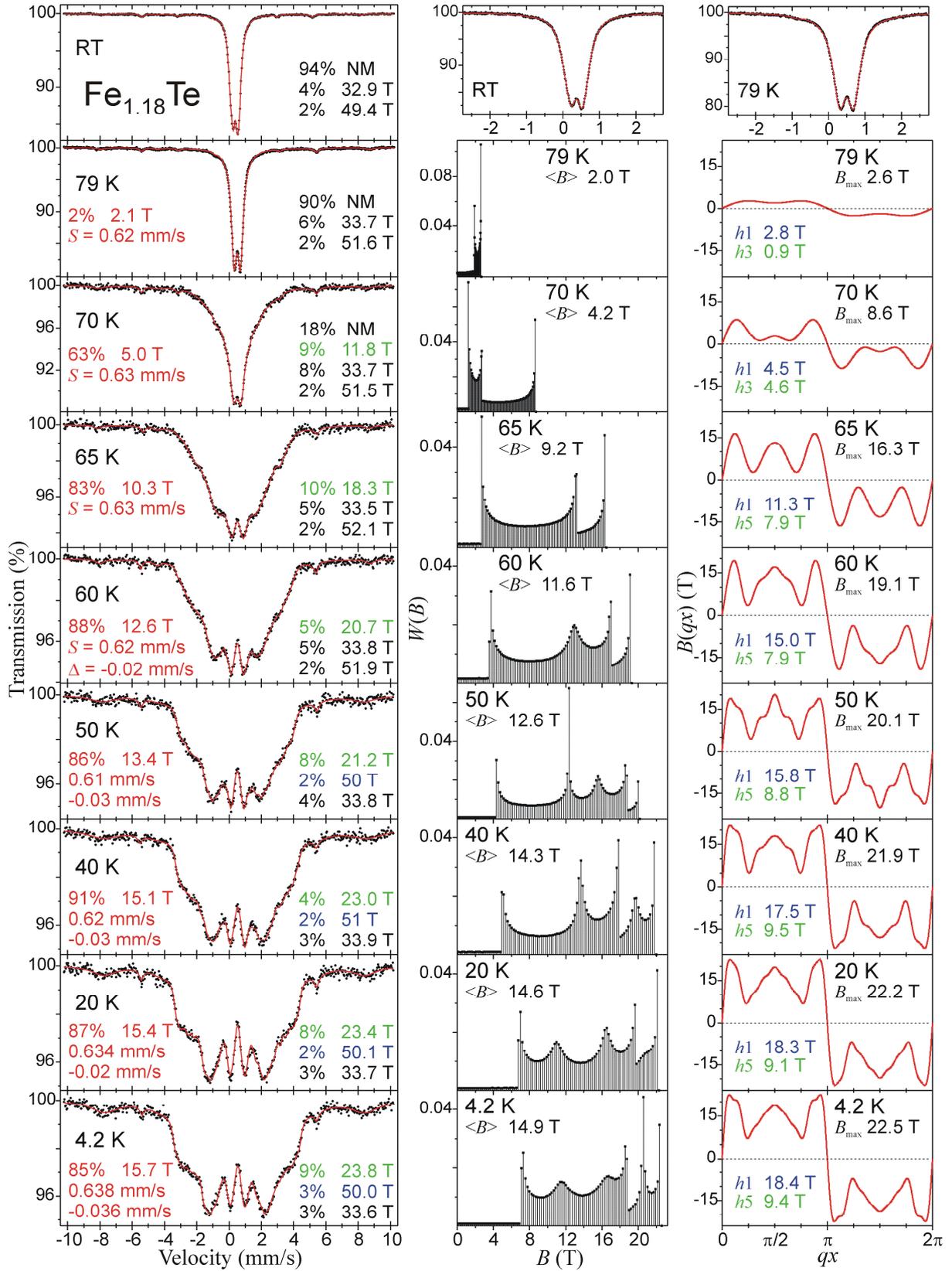

Fig. 4 $^{57}$Fe Mössbauer spectra, hyperfine field distributions of SDW and SDW shape versus temperature for Fe$_{1.18}$Te. See text for details.

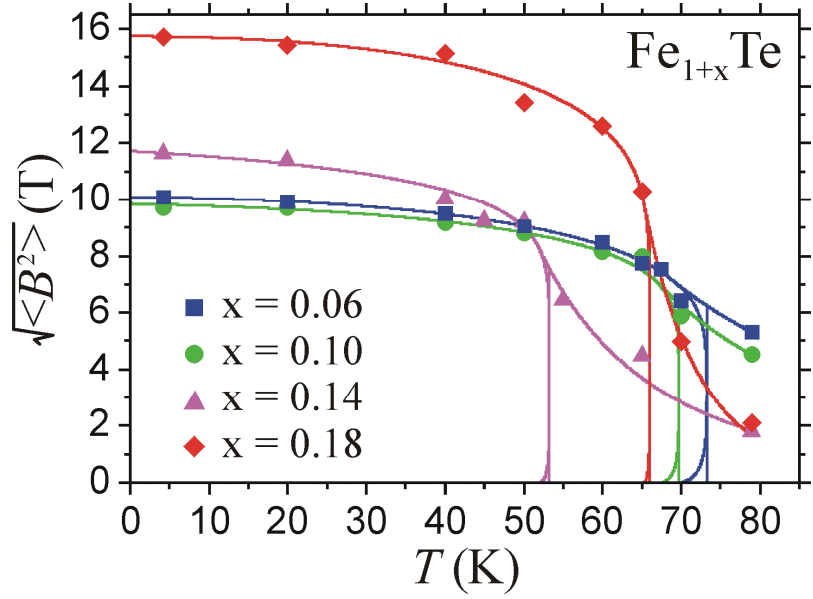

Fig. 5 Root mean square amplitude of the SDW versus temperature for $Fe_{1+x}Te$. Experimental errors do not exceed size of the symbol used to mark the experimental data. Solid lines represent total, coherent and incoherent contributions [18]. The coherent contribution falls to null at temperature $T_c$, while the incoherent part appears at and just above lower temperature $T_0$.